\documentclass{article}%
\usepackage{amsfonts}
\usepackage{amsmath}%
\usepackage{amssymb}%
\usepackage{graphicx}
\newtheorem{theorem}{Theorem}
\newtheorem{acknowledgement}[theorem]{Acknowledgement}

\newtheorem{conclusion}[theorem]{Conclusion}

\newtheorem{definition}[theorem]{Definition}

\newtheorem{remark}[theorem]{Remark}

\begin{document}

\title{Quantum Algorithm for SAT Problem and Quantum Mutual Entropy}
\author{Masanori Ohya\\Department of Information Sciences, \\Tokyo University of Science, \\278 Noda City, Chiba, Japan }
\maketitle

\begin{abstract}
It is von Neumann who opened the window for today's Information epoch. He
defined quantum entropy including Shannon's information more than 20 years
ahead of Shannon, and he introduced a concept what computation means mathematically.

In this paper I will report two works that we have recently done, one of which
is on quantum algorithum in generalized sense solvimg the SAT problem (one of
NP complete problems) and another is on quantum mutual entropy properly
describing quantum communication processes.

\end{abstract}

\section{Introduction}

This paper consists of two parts, one (Sec.2 and 3) of which is about quantum
algorithm solving the SAT problem based on a series of the papers
\cite{OM,AS,OV1,OV2,AO2} and another (Sec.4) is about quantum mutual entropy
applying quantum communication processses based on the papers \cite{O3,O4,OW}.

Although the ability of computer is highly progressed, there are several
problems which may not be solved effectively, namely, in polynomial time.
Among such problems, NP problem and NP complete problem are fundamental. It is
known that all NP complete (NPC for short) problems are equivalent and have
been studied for decades, for which all known algorithms have an exponential
running time in the length of the input so far.. An essential question to be
asked for more than 30 years is\textit{\ whether there exists an algorithm to
solve an NP complete problem in polynomial time}. We found two different
algorithms solving the NPC problems in polynomial time \cite{OM,OV1,OV2,AO}.
In first two sections of the present paper we report the essence of these algorithms.

After von Neumann introduced quantum entropy \cite{Neu} of density operators,
many studies on various quantum entropies have appeared \cite{Neu,OP},among
which quantum mutual entropy plays an important role. That is, the mutual
entropy expresses the amount of information sending from input to output, so
that it will be a basic quantity measuring the ability of a communication
channel. I defined the quantum mutual entropy for desity operators in 1983
\cite{O3} by using Umegaki's relative entropy \cite{Um} and extended it to
general C*-dynamical systems by means of Araki's or Uhlmann's relative entropy
\cite{Ara,Uhl,OP}. Recently several quantum mutual type entropies have
appeared \cite{Sho2,BNS,BSST}, and they are used to discuss communication
processes. In Section 4 of this paper, we compare these mutual type entropies
from the views of information communication based on the paper \cite{OW}.

\section{Quantum Chaos Algorithm of SAT}

Let us remind what the P-problem and the NP-problem are \cite{GJ,Cle}: Let $n
$ be the size of input. \newline

(1)A P-problem is a problem whose time needed for solving the problem is at
worst of polynomial time of $n.$ Equivalently, it is a problem which can be
recognized in a polynomial time of $n$ by deterministic Turing machine.
\newline

(2)An NP-problem is a problem that can be solved in polynomial time by a
nondeterministic Turing machine. This can be understood as follows: Let
consider a problem to find a solution of $f\left(  x\right)  =0$. We can check
in\ polynomial time of $n$ whether $x_{0}$ is a solution of $f\left(
x\right)  =0$, but we do not\ know whether we can find the solution of
$f\left(  x\right)  =0$ in polynomial time of $n$.

(3) An NP-complete problem is a problem polynomialy transformed NP-problem.

We take the SAT (satisfiable) problem, one of the NP-complete problems, to
study whether there exists an algorithm showing NPC becomes P. It is known
that the SAT problem is equivalent to any other NPC problems.

Let $X\equiv\left\{  {x_{1},\cdots,x_{n}}\right\}  $ be a set. Then $x_{k}$
and its negation $\bar{x}_{k}\left(  {k=1,2,\cdots,n}\right)  $ are called
literals and the set of all such literals is denoted by $\widetilde{X}%
\equiv\left\{  {x_{1},\bar{x}_{1},\cdots,x_{n},\bar{x}_{n}}\right\}  $. The
set of all subsets of $X^{\prime}$ is denoted by $\mathcal{F}\left(
\widetilde{X}\right)  $ and an element $C\in\mathcal{F}\left(  \widetilde
{X}\right)  $ is called a clause. We take a truth assignment to all Boolean
variables $x_{k}$. If we can assign the truth value to at least one element of
$C$, then $C$ is called satisfiable. When $C$ is satisfiable, the truth value
$t\left(  C\right)  $ of $C$ is regarded as true, otherwise, that of $C$ is
false. Take the truth values as \textquotedblright true $\leftrightarrow$1,
false $\leftrightarrow$0\textquotedblright. Then $C$is satisfiable iff
$t\left(  C\right)  =1$.

Let $L=\left\{  {0,1}\right\}  $ be a Boolean lattice with usual join $\vee$
and meet $\wedge$, and $t\left(  x\right)  $ be the truth value of a literal
$x$ in $X$. Then the truth value of a clause $C$ is written as $t\left(
C\right)  \equiv\vee_{x\in C}t\left(  x\right)  $.

Moreover the set $\mathcal{C}$ of all clauses $C_{j}\left(  {j=1,2,\cdots
,m}\right)  $ is called satisfiable iff the meet of all truth values of
$C_{j}$ is 1; $t\left(  \mathcal{C}\right)  \equiv\wedge_{j=1}^{m}t\left(
{C_{j}}\right)  =1$. Thus the SAT problem is written as follows:

\begin{definition}
SAT Problem: Given a Boolean set $X\equiv\left\{  {x_{1},\cdots,x_{n}%
}\right\}  $and a set $\mathcal{C}=\left\{  \mathcal{C}_{1},\cdots
,\mathcal{C}_{m} \right\}  $ of clauses, determine whether $\mathcal{C}$ is
satisfiable or not.
\end{definition}

That is, this problem is to ask whether there exists a truth assignment to
make $\mathcal{C}$ satisfiable. It is known in usual algorithm that it is
polynomial time to check the satisfiability only when a specific truth
assignment is given, but we can not determine the satisfiability in polynomial
time when an assignment is not specified.

In \cite{OM} we discussed the quantum algorithm of the SAT problem, which was
rewritten in \cite{AS} with showing that the OM SAT-algorithm is combinatric.
Ohya and Masuda pointed out \cite{OM} that the SAT problem, hence all other NP
problems, can be solved in polynomial time by quantum computer if the
superposition of two orthogonal vectors $\left\vert 0\right\rangle $ and
$\left\vert 1\right\rangle $ is physically detected. However this detection is
considered not to be possible in the present technology. The problem to be
overcome is how to distinguish the pure vector $\left\vert 0\right\rangle $
from the superposed one $\alpha\left\vert 0\right\rangle +\beta\left\vert
1\right\rangle ,$ obtained by the OM SAT-quantum algorithm, if $\beta$ is not
zero but very small. If such a distinction is possible, then we can solve the
NPC problem in the polynomial time.

In \cite{OV1,OV2} it is shown that it can be possible by combining nonlinear
chaos amplifier with the quantum algorithm, which implies the existence of a
mathematical algorithm solving NP=P. The algorithm of Ohya and Volovich is
going beyond usual (unitary) quantum Turing algorithm. So the next question is
(1) whether there exists more general Turing machine scheme combining the
unitary quantum algorithm with chaos dynamics, or (2) whether there exists
another method to achieve the above distinction of two vectors by a suitable
unitary evolution. In the paper \cite{AO2}, we discussed that the stochastic
limit, recently extensively studied by Accardi and coworkers \cite{ALV}, can
be used to find another method of (2).

In this paper, we review mathematical frame of quantum algorithm in Section 2
and the OV-chaos algorithm. In Section 3, based on the idea of quantum
adaptive dynamics \cite{AI,KOT,AO2}, we discuss how it can be used to solve
the problem NP=P.

\subsection{Quantum algorithm}

The quantum algorithms discussed so far are rather idealized because
computation is represented by unitary operations. A unitary operation is
rather difficult to realize in physical processes, more realistic operation is
one allowing some dissipation like semigroup dynamics. For such a realization,
we have to generalize the concept of quantum Turing machine so that the
generalized one contains non-unitary operations. This work has been done in
the papers \cite{AO2,IOV,AO3}, about which we will not discuused here. We
will, in this paper, explain the algorithms solving the SAT problem in
polynominal time.

First we remind the precedure of usual quantum algorithm which is needed the
computation of the truth value $t\left(  \mathcal{C}\right)  $ of the SAT.

Let $\mathcal{H}$ be a Hilbert space describing input, computation and output
(result). As usual, the Hilbert space is $\mathcal{H}=\otimes_{1}%
^{N}\mathbf{C}^{2}$, and let the basis of $\mathcal{H}=\otimes_{1}%
^{N}\mathbf{C}^{2}$ be: $e_{0}\left(  {=\left|  0\right\rangle }\right)
=\left|  0\right\rangle \otimes\cdots\otimes\left|  0\right\rangle
\otimes\left|  0\right\rangle ,e_{1}\left(  {=\left|  1\right\rangle }\right)
=\left|  0\right\rangle \otimes\cdots\otimes\left|  0\right\rangle
\otimes\left|  1\right\rangle ,\cdots,e_{2^{N}-1}\left(  {=\left|  {2^{N}%
-1}\right\rangle }\right)  =\left|  1\right\rangle \otimes\cdots\otimes\left|
1\right\rangle \otimes\left|  1\right\rangle .$

Any number $t$ $\left(  {0,\cdots,2^{N}-1}\right)  $ can be expressed by
$t=\sum\limits_{k=1}^{N}{a_{t}^{\left(  k\right)  }}2^{k-1},$ ${a_{t}^{\left(
k\right)  }=0}$ ${or}$ 1, so that the associated vector is written by%

\[
\left\vert t\right\rangle \left(  {=e_{t}}\right)  =\otimes_{k=1}%
^{N}\left\vert {a_{t}^{\left(  k\right)  }}\right\rangle .
\]
And applying n-tuples of Hadamrd matrix $H\equiv\frac{1}{\sqrt{2}}\left(
\begin{array}
[c]{cc}%
1 & 1\\
1 & -1
\end{array}
\right)  $ to the vacuum vector $\left\vert 0\right\rangle ,$ we get
$H\left\vert 0\right\rangle \left(  {\ =\xi\left(  0\right)  }\right)
\equiv\otimes_{1}^{N}\frac{1}{\sqrt{2}}\left(  {\left\vert 0\right\rangle
+\left\vert 1\right\rangle }\right)  .$ Put%

\[
W\left(  t\right)  \equiv\otimes_{j=1}^{N}\left(
\begin{array}
[c]{cc}%
1 & 0\\
0 & \exp(\frac{2\pi it}{2^{N}}2^{j-1})
\end{array}
\right)  .
\]
Then we have%

\[
\xi\left(  t\right)  \equiv W\left(  t\right)  \xi\left(  0\right)  =\frac
{1}{\sqrt{2^{N}}}\sum\limits_{k=0}^{2^{N}-1}{exp\left(  {\frac{{2\pi itk}%
}{{2^{N}}}}\right)  }\left\vert k\right\rangle ,
\]
which is called Discrete Fourier Transformation. Thus altogether of the above
operations, it follows a unitary operator $U_{F}\left(  t\right)  \equiv
W\left(  t\right)  H$ and the vector $\xi\left(  t\right)  =U_{F}\left(
t\right)  \left\vert 0\right\rangle .$

All conventional unitary algorithms can be written as the following three
steps by means of certain channels on the state space in $\mathcal{H}$ (i.e.,
a channel is a map sending a state to another state) :

(1) Preparation of state: Take a state $\rho$ (e.g., $\rho=\left|
0\right\rangle \left\langle 0\right|  $) applying the unitary channel defined
by the above $U_{F}\left(  t\right)  :\Lambda_{F}^{\ast}=Ad_{U_{F}\left(
t\right)  }$%
\[
\Lambda_{F}^{\ast}=Ad_{U_{F}}\Longrightarrow\Lambda_{F}^{\ast}\rho=U_{F}\rho
U_{F}^{\ast}%
\]

(2) Computation: Let $U$ a unitary operator on $\mathcal{H}$ representing the
computation followed by a suitable programming of a certain problem, then the
computation is described by a channel $\Lambda_{U}^{\ast}=Ad_{U}$ (unitary
channel). After the computation, the final state $\rho_{f}$ will be%

\[
\rho_{f}={\Lambda_{U}^{\ast}\Lambda_{F}^{\ast}\rho.}%
\]

(3) Register and Measurement: For registeration of the computed result and its
measurement we might need an additional system $\mathcal{K}$ (e.g., register),
so that the lifting $\mathcal{E}_{m}^{\ast}$ from $\mathcal{S}\left(
\mathcal{H}\right)  $ to $\mathcal{S}\left(  {\mathcal{H}\otimes\mathcal{K}%
}\right)  $ in the sense of \cite{AO} is useful to describe this stage. Thus
the whole process is wrtten as$\ \ \ \ \ $%

\[
\rho_{f}=\mathcal{E}_{m}^{\ast}\left(  {\Lambda_{U}^{\ast}\Lambda_{F}^{\ast
}\rho}\right)  .
\]
Finally we measure the state in $\mathcal{K}$: For instance, let $\left\{
{P_{k};k\in J}\right\}  $ be a projection valued measure (PVM) on
$\mathcal{K}$%
\[
\Lambda_{m}^{\ast}\rho_{f}=\sum\limits_{k\in J}I\otimes{P_{k}}\rho_{f}I\otimes
P_{k},
\]
after which we can get a desired result by observations in finite times if the
size of the set $J$ is small.

\begin{remark}
When dissipation is involved the above three steps have to be generalized so
that dissipative nature is involved. Such a generalization can be expressed by
means of suitable channel, not necessarily unitary.(1) Preparation of state:
We may be use the same channel $\Lambda_{F}^{\ast}=Ad_{U_{F}}$ in this first
step, but if the number of qubits $N$ is large so that it will not be built
physically, then $\Lambda_{F}^{\ast}\ $should be modified, and let denote it
by $\Lambda_{P}^{\ast}.$(2) Computation: This stage is certainly modified to a
channel $\Lambda_{C}^{\ast}$ reflecting the physical device for computer.(3)
Registering and Measurement: This stage will be remained as aobe. Thus the
whole process is written as$\ \ \ \ \ $%
\[
\rho_{f}=\mathcal{E}_{m}^{\ast}\left(  {\Lambda_{C}^{\ast}\Lambda_{P}^{\ast
}\rho}\right)  .
\]

\end{remark}

\subsection{Quantum algorithm of SAT}

We explain the algorithm of the SAT problem which has been introduced by
Ohya-Masuda \cite{OM} and developed by Accardi-Sabbadini \cite{AS}. This
quantum algorithm is described by a combination of the unitary operators
discussed in the previous section on a Hilbert space $\mathcal{H}$. The detail
of this section is given in the papers \cite{OM,AS,OV2}, so we will discuss
just the essence of the OM algorithm. Throughout this subsection, let $n$ be
the total number of Boolean variables used in the SAT problem.

\bigskip Let 0 and 1 of the Boolean lattice $L$ be denoted by the vectors
$\left\vert 0\right\rangle \equiv\left(
\begin{array}
[c]{l}%
1\\
0
\end{array}
\right)  $ and $\left\vert 1\right\rangle \equiv\left(
\begin{array}
[c]{l}%
0\\
1
\end{array}
\right)  $ in the Hilbert space \textbf{C}$^{2},$ respectively. That is, the
vector $\left\vert 0\right\rangle $ corresponds to falseness and $\left\vert
1\right\rangle $ does to truth.

As we explained in the previous section, an element $x\in X$ can be denoted by
0 or 1, so by $\left|  0\right\rangle $ or $\left|  1\right\rangle .$ In order
to describe a clause $C$ with at most $n$ length by a quantum state, we need
the n-tuple tensor product Hilbert space $\mathcal{H\equiv}$ $\otimes_{1}^{n}%
$\textbf{C}$^{2}.$ For instance, in the case of $n=2$, \ given $C=\left\{
x_{1},x_{2}\right\}  $ with an assignment $x_{1}=0$ and $x_{2}=1,$ then the
corresponding quantum state vector is $\left|  0\right\rangle \otimes\left|
1\right\rangle ,$ so that the quantum state vector describing $C$ is generally
written by $\left|  C\right\rangle =\left|  x_{1}\right\rangle \otimes\left|
x_{2}\right\rangle \in$ $\mathcal{H}$ with $x_{k}=0$ or $1$ (k=1,2).

Once $X\equiv\left\{  x_{1},\cdots,x_{n}\right\}  $ and $\mathcal{C=}\left\{
C_{1},C_{2},\cdots,C_{m}\right\}  $ are given, the SAT is to find the vector%

\[
\left\vert t\left(  \mathcal{C}\right)  \right\rangle \equiv\wedge_{j=1}%
^{m}\vee_{x\in C_{j}}t(x),
\]
where $t(x)$ is $\left\vert 0\right\rangle \ $or $\left\vert 1\right\rangle $
when $x=0$ or 1, respectively, and $t(x)\wedge t(y)\equiv t(x\wedge y)$,
$t(x)\vee t(y)\equiv t(x\vee y).$

For any two qubits $\left\vert x\right\rangle $ and $\left\vert y\right\rangle
$, $\left\vert x,y\right\rangle $ and $\left\vert x^{N}\right\rangle $ is
defined as $\left\vert x\right\rangle \otimes\left\vert y\right\rangle $ and
$\underset{N\text{ times}}{\underbrace{\left\vert x\right\rangle \otimes
\cdots\otimes\left\vert x\right\rangle }} $, respectively. The usual (unitary)
quantum computation can be formulated mathematically as the multiplication by
unitary operators. Let $U_{NOT}$,$U_{CN}$ and $U_{CCN}$ be the three unitary
operators defined as%
\begin{align*}
U_{NOT} &  \equiv\left\vert 1\right\rangle \left\langle 0\right\vert
+\left\vert 0\right\rangle \left\langle 1\right\vert ,\\
U_{CN} &  \equiv\left\vert 0\right\rangle \left\langle 0\right\vert \otimes
I+\left\vert 1\right\rangle \left\langle 1\right\vert \otimes U_{NOT},\\
U_{CCN} &  \equiv\left\vert 0\right\rangle \left\langle 0\right\vert \otimes
I\otimes I+\left\vert 1\right\rangle \left\langle 1\right\vert \otimes
\left\vert 0\right\rangle \left\langle 0\right\vert \otimes I+\left\vert
1\right\rangle \left\langle 1\right\vert \otimes\left\vert 1\right\rangle
\left\langle 1\right\vert \otimes U_{NOT}.
\end{align*}
$U_{NOT}$,$U_{CN}$ and $U_{CCN}$ are often called NOT-gate, Controlled-NOT
gate and Controlled-Controlled-NOT gate, respectively. For any $k\in
\mathbb{N}$, $U_{H}^{\left(  N\right)  }\left(  k\right)  $ denotes the
$k$-tuple Hadamard transformation on $\left(  \mathbb{C}^{2}\right)  ^{\otimes
N}$ defined as%

\[
U_{H}^{\left(  N\right)  }\left(  k\right)  \left\vert 0^{N}\right\rangle
=\frac{1}{2^{k/2}}\left(  \left\vert 0\right\rangle +\left\vert 1\right\rangle
\right)  ^{\otimes k}\left\vert 0^{N-k}\right\rangle =\frac{1}{2^{k/2}}%
\sum\limits_{i=0}^{2^{k-1}}\left\vert e_{i}\right\rangle \otimes\left\vert
0^{N-k}\right\rangle .
\]

The above unitary operators can be extended to the unitary operators on
$\left(  \mathbb{C}^{2}\right)  ^{\otimes N}$:%

\begin{align*}
U_{NOT}^{\left(  N\right)  }(u) &  \equiv I^{\otimes u-1}\otimes\left(
\left\vert 0\right\rangle \left\langle 1\right\vert +\left\vert 1\right\rangle
\left\langle 0\right\vert \right)  I^{\otimes N-u-1}\\
U_{CN}^{\left(  N\right)  }\left(  u,v\right)   &  \equiv I^{\otimes
u-1}\otimes\left\vert 0\right\rangle \left\langle 0\right\vert \otimes
I^{\otimes N-u-1}+I^{\otimes u-1}\otimes\left\vert 1\right\rangle \left\langle
1\right\vert \otimes\\
&  \text{ }I^{\otimes v-u-1}\otimes U_{NOT}\otimes I^{\otimes N-v-1}\\
U_{CCN}^{\left(  N\right)  }\left(  u,v,w\right)   &  =I^{\otimes u-1}%
\otimes\left\vert 0\right\rangle \left\langle 0\right\vert \otimes I^{\otimes
N-u-1}+I^{\otimes u-1}\otimes\left\vert 1\right\rangle \left\langle
1\right\vert \otimes\\
&  I^{\otimes v-u-1}\otimes\left\vert 0\right\rangle \left\langle 0\right\vert
\otimes I^{\otimes N-v-1}+I^{\otimes u-1}\otimes\left\vert 1\right\rangle
\left\langle 1\right\vert \otimes\\
&  \text{ }I^{\otimes v-u-1}\otimes\left\vert 1\right\rangle \left\langle
1\right\vert \otimes I^{\otimes w-t-1}\otimes U_{NOT}\otimes I^{\otimes
N-w-1},
\end{align*}
where $u,v$ and $w$ be positive integers satisfying $1\leq u<v<w\leq N$.

Furthermore we have the following three unitary operators $U_{AND},U_{OR}$ and
$U_{COPY}$ , called the logical gates; (see \cite{AS})%

\begin{align*}
U_{AND}  &  \equiv\sum_{\varepsilon_{1},\varepsilon_{2}\in\left\{
0,1\right\}  }\left\{  \left\vert \varepsilon_{1},\varepsilon_{2}%
,\varepsilon_{1}\wedge\varepsilon_{2}\right\rangle \left\langle \varepsilon
_{1},\varepsilon_{2},0\right\vert +\left\vert \varepsilon_{1},\varepsilon
_{2},1-\varepsilon_{1}\wedge\varepsilon_{2}\right\rangle \left\langle
\varepsilon_{1},\varepsilon_{2},1\right\vert \right\} \\
&  =\left\vert 0,0,0\right\rangle \left\langle 0,0,0\right\vert +\left\vert
0,0,1\right\rangle \left\langle 0,0,1\right\vert +\left\vert
1,0,0\right\rangle \left\langle 1,0,0\right\vert +\left\vert
1,0,1\right\rangle \left\langle 1,0,1\right\vert \\
&  +\left\vert 0,1,0\right\rangle \left\langle 0,1,0\right\vert +\left\vert
0,1,1\right\rangle \left\langle 0,1,1\right\vert +\left\vert
1,1,1\right\rangle \left\langle 1,1,0\right\vert +\left\vert
1,1,0\right\rangle \left\langle 1,1,1\right\vert .
\end{align*}

\begin{align*}
U_{OR}  &  \equiv\sum_{\varepsilon_{1},\varepsilon_{2}\in\left\{  0,1\right\}
}\left\{  \left\vert \varepsilon_{1},\varepsilon_{2},\varepsilon_{1}%
\vee\varepsilon_{2}\right\rangle \left\langle \varepsilon_{1},\varepsilon
_{2},0\right\vert +\left\vert \varepsilon_{1},\varepsilon_{2},1-\varepsilon
_{1}\vee\varepsilon_{2}\right\rangle \left\langle \varepsilon_{1}%
,\varepsilon_{2},1\right\vert \right\} \\
&  =\left\vert 0,0,0\right\rangle \left\langle 0,0,0\right\vert +\left\vert
0,0,1\right\rangle \left\langle 0,0,1\right\vert +\left\vert
1,0,1\right\rangle \left\langle 1,0,0\right\vert +\left\vert
1,0,0\right\rangle \left\langle 1,0,1\right\vert \\
&  +\left\vert 0,1,1\right\rangle \left\langle 0,1,0\right\vert +\left\vert
0,1,0\right\rangle \left\langle 0,1,1\right\vert +\left\vert
1,1,1\right\rangle \left\langle 1,1,0\right\vert +\left\vert
1,1,0\right\rangle \left\langle 1,1,1\right\vert .
\end{align*}

\begin{align*}
U_{COPY} &  \equiv\sum_{\varepsilon_{1}\in\left\{  0,1\right\}  }\left\{
\left\vert \varepsilon_{1},\varepsilon_{1}\right\rangle \left\langle
\varepsilon_{1},0\right\vert +\left\vert \varepsilon_{1},1-\varepsilon
_{1}\right\rangle \left\langle \varepsilon_{1},1\right\vert \right\} \\
&  =\left\vert 0,0\right\rangle \left\langle 0,0\right\vert +\left\vert
0,1\right\rangle \left\langle 0,1\right\vert +\left\vert 1,1\right\rangle
\left\langle 1,0\right\vert +\left\vert 1,0\right\rangle \left\langle
1,1\right\vert .
\end{align*}
Here $\varepsilon_{1}$ and $\varepsilon_{2}$ take the value 0 or 1. We call
$U_{AND},U_{OR}$ and $U_{COPY}$ , AND gate, OR gate and COPY gate,
respectively, whose extensions to $\left(  \mathbb{C}^{2}\right)  ^{\otimes
N}$ are denoted by $U_{AND}^{\left(  N\right)  },U_{OR}^{\left(  N\right)  }$
and $U_{COPY}^{\left(  N\right)  }$, which are expressed as%

\begin{align*}
U_{AND}^{\left(  N\right)  }(u,v,w)  &  =\sum_{\varepsilon_{1},\varepsilon
_{2}\in\left\{  0,1\right\}  }I^{\otimes u-1}\otimes\left\vert \varepsilon
_{1}\right\rangle \left\langle \varepsilon_{1}\right\vert I^{\otimes
v-u-1}\otimes\left\vert \varepsilon_{2}\right\rangle \left\langle
\varepsilon_{2}\right\vert \\
&  I^{\otimes w-v-u-1}\otimes\left\vert \varepsilon_{1}\wedge\varepsilon
_{2}\right\rangle \left\langle 0\right\vert I^{\otimes N-w-v-u}+\\
&  I^{\otimes u-1}\otimes\left\vert \varepsilon_{1}\right\rangle \left\langle
\varepsilon_{1}\right\vert I^{\otimes v-u-1}\otimes\\
&  \left\vert \varepsilon_{2}\right\rangle \left\langle \varepsilon
_{2}\right\vert I^{\otimes w-v-u-1}\otimes\left\vert 1-\varepsilon_{1}%
\wedge\varepsilon_{2}\right\rangle \left\langle 1\right\vert I^{\otimes
N-w-v-u}.
\end{align*}%
\begin{align*}
U_{OR}^{\left(  N\right)  }\left(  u,v,w\right)   &  \equiv\sum_{\varepsilon
_{1},\varepsilon_{2}\in\left\{  0,1\right\}  }I^{\otimes u-1}\otimes\left\vert
\varepsilon_{1}\right\rangle \left\langle \varepsilon_{1}\right\vert
I^{\otimes v-u-1}\otimes\left\vert \varepsilon_{2}\right\rangle \left\langle
\varepsilon_{2}\right\vert \\
&  I^{\otimes w-v-u-1}\otimes\left\vert \varepsilon_{1}\vee\varepsilon
_{2}\right\rangle \left\langle 0\right\vert I^{\otimes N-w-v-u}+\\
&  I^{\otimes u-1}\otimes\left\vert \varepsilon_{1}\right\rangle \left\langle
\varepsilon_{1}\right\vert I^{\otimes v-u-1}\otimes\left\vert \varepsilon
_{2}\right\rangle \left\langle \varepsilon_{2}\right\vert \\
&  I^{\otimes w-v-u-1}\otimes\left\vert 1-\varepsilon_{1}\vee\varepsilon
_{2}\right\rangle \left\langle 1\right\vert I^{\otimes N-w-v-u}.
\end{align*}%
\begin{align*}
U_{COPY}^{\left(  N\right)  }\left(  u,v\right)   &  \equiv\sum_{\varepsilon
_{1}\in\left\{  0,1\right\}  }I^{\otimes u-1}\left\vert \varepsilon
_{1}\right\rangle \left\langle \varepsilon_{1}\right\vert I^{\otimes
v-u-1}\left\vert \varepsilon_{1}\right\rangle \left\langle 0\right\vert
I^{\otimes N-v-u}\\
&  +I^{\otimes u-1}\left\vert \varepsilon_{1}\right\rangle \left\langle
\varepsilon_{1}\right\vert I^{\otimes v-u-1}\left\vert 1-\varepsilon
_{1}\right\rangle \left\langle 1\right\vert I^{\otimes N-v-u}.
\end{align*}
where $u,v$ and $w$ are positive integers satisfying $1\leq u<v<w\leq N$.
These operators can be written, in terms of elementary gates, as%
\begin{align*}
U_{OR}^{\left(  N\right)  }\left(  u,v,w\right)   &  =U_{CN}^{\left(
N\right)  }\left(  u,w\right)  \cdot U_{CN}^{\left(  N\right)  }\left(
v,w\right)  \cdot U_{CCN}^{\left(  N\right)  }\left(  u,v,w\right)  ,\\
U_{AND}^{\left(  N\right)  }\left(  u,v,w\right)   &  =U_{CCN}^{\left(
N\right)  }\left(  u,v,w\right)  ,\\
U_{COPY}^{\left(  N\right)  }\left(  u,v\right)   &  =U_{CN}^{\left(
N\right)  }\left(  u,v\right)  .
\end{align*}

Let $\mathcal{C}$ be a set of clauses whose cardinality is equal to $m$. Let
$\mathcal{H}=\left(  \mathbf{C}^{2}\right)  ^{\otimes n+\mu+1}$ be a Hilbert
space and $\left\vert v_{0}\right\rangle $ be the initial state $\left\vert
v_{0}\right\rangle =\left\vert 0^{n},0^{\mu},0\right\rangle $, where $\mu$ is
the number of dust qubits (the details are seen in \cite{IA}). Let
$U_{\mathcal{C}}^{\left(  n\right)  }$ be a unitary operator for the
computation of the SAT:%

\[
U_{\mathcal{C}}^{\left(  n\right)  }\left\vert v_{0}\right\rangle =\frac
{1}{\sqrt{2^{n}}}\sum_{i=0}^{2^{n}-1}\left\vert e_{i},x^{\mu},t_{e_{i}}\left(
\mathcal{C}\right)  \right\rangle \equiv\left\vert v_{f}\right\rangle
\]
where $x^{\mu}$ denotes a $\mu$ strings of binary symbols and $t_{e_{i}%
}\left(  \mathcal{C}\right)  $ is a truth value of $\mathcal{C}$ with $e_{i}$.

Let $\left\{  s_{k};k=1,\dots,m\right\}  $ be the sequence defined as%
\begin{align*}
s_{1} &  =n+1,\\
s_{2} &  =s_{1}+card\left(  C_{1}\right)  +\delta_{1,card\left(  C_{1}\right)
}-1,\\
s_{i} &  =s_{i-1}+card\left(  C_{i-1}\right)  +\delta_{1,card\left(
C_{i-1}\right)  },\text{ \ \ }3\leq i\leq m,
\end{align*}
where $card\left(  C_{i}\right)  $ means the cardinality of a clause $C_{i}$.
Take a value $s$ as%
\[
s=s_{m}-1+card\left(  C_{m}\right)  +\delta_{1,card\left(  C_{m}\right)  }.
\]
Note that the number $m$ of the clause is at most $2n$. Then we have
\cite{IA}: The total number of dust qubits $\mu$ is
\begin{align*}
\mu &  =s-1-n\\
&  =\sum_{k=1}^{m}card\left(  C_{k}\right)  +\delta_{1,card\left(
C_{k}\right)  }-2
\end{align*}
for $m\geq2.$ In order to construct $U_{\mathcal{C}}^{\left(  n\right)  }$
concretely, we use the following unitary gates for this concrete expression
\cite{OM,AS}:%

\[
U_{AND}^{\left(  x\right)  }\left(  k\right)  =\left\{
\begin{array}
[c]{c}%
U_{AND}^{\left(  x\right)  }\left(  s_{k+1}-1,s_{k+2}-2,s_{k+2}-1\right)
,\text{ \ \ \ }1\leq k\leq m-2\\
U_{AND}^{\left(  x\right)  }\left(  s_{m}-1,s_{f}-1,s_{f}\right)  ,\text{
\ \ \ \ }k=m-1
\end{array}
\right.  ,
\]

\begin{align*}
U_{OR}^{\left(  x\right)  }\left(  k\right)   &  =\bar{U}_{OR}^{\left(
x\right)  }\left(  l_{4},s_{k}-card\left(  C_{k}\right)  -1,s_{k}-card\left(
C_{k}\right)  -2\right)  \cdot\cdots\cdot\bar{U}_{OR}^{\left(  x\right)
}\left(  l_{3},s_{k},s_{k}+1\right)  \bar{U}_{OR}^{\left(  x\right)  }\left(
l_{1},l_{2},s_{k}\right)  ,\\
\bar{U}_{OR}^{\left(  x\right)  }\left(  u,v,w\right)   &  =\left\{
\begin{array}
[c]{c}%
U_{OR}^{\left(  x\right)  }\left(  u,v,w\right)  ,\text{ \ \ \ \ \ \ \ \ }%
x_{u}\in C_{k}\\
U_{NOT}^{\left(  x\right)  }\left(  u\right)  \cdot U_{OR}^{\left(  x\right)
}\left(  u,v,w\right)  \cdot U_{NOT}^{\left(  x\right)  }\left(  u\right)
,\text{ \ \ \ \ \ }\bar{x}_{u}\in C_{k}\\
U_{NOT}^{\left(  x\right)  }\left(  u\right)  \cdot U_{NOT}^{\left(  x\right)
}\left(  v\right)  \cdot U_{OR}^{\left(  x\right)  }\left(  u,v,w\right)
\cdot U_{NOT}^{\left(  x\right)  }\left(  u\right)  \cdot U_{NOT}^{\left(
x\right)  }\left(  v\right)  ,\text{ \ \ \ \ \ }\bar{x}_{u},\bar{x}_{v}\in
C_{k}%
\end{array}
\right.  ,
\end{align*}
where $l_{1},l_{2},l_{3},l_{4}$ are positive integers such that $x_{z}\in
C_{k}$ or $\bar{x}_{z}\in C_{k}$, $\left(  z=l_{1},\ldots,l_{4}\right)  $.

\begin{theorem}
The unitary operator $U_{\mathcal{C}}^{\left(  n\right)  }$, is represented as%
\begin{align*}
U_{\mathcal{C}}^{\left(  n\right)  } &  =U_{AND}^{\left(  n+\mu+1\right)
}\left(  m-1\right)  \cdot U_{AND}^{\left(  n+\mu+1\right)  }\left(
m-2\right)  \cdot\cdots\cdot U_{AND}^{\left(  n+\mu+1\right)  }\left(
1\right) \\
&  \cdot U_{OR}^{\left(  n+\mu+1\right)  }\left(  m\right)  \cdot
U_{OR}^{\left(  n+\mu+1\right)  }\left(  m-1\right)  \cdot\cdots\cdot
U_{OR}^{\left(  n+\mu+1\right)  }\left(  1\right)  \cdot U_{H}^{\left(
n+\mu+1\right)  }\left(  n\right)  .
\end{align*}

\end{theorem}

Applying the aboved unitary operator to the initial state, we obtain the final
state $\rho.$The result of the computation is registered as $\left\vert
t\left(  \mathcal{C}\right)  \right\rangle $ in the last section of the final
vector, which will be taken out by a projection $P_{n+\mu,1}\equiv I^{\otimes
n+\mu}\otimes\left\vert 1\right\rangle \left\langle 1\right\vert $ onto the
subspace of $\mathcal{H}$ spanned by the vectors $\left\vert \varepsilon
^{n},\varepsilon^{\mu},1\right\rangle $.

The following theorem is easily seen.

\begin{theorem}
$\mathcal{C}$ is SAT if and only if%
\[
P_{n+\mu,1}U_{\mathcal{C}}^{\left(  n\right)  }\left\vert v_{0}\right\rangle
\neq0
\]

\end{theorem}

According to the standard theory of quantum measurement, after a measurement
of the event $P_{n+\mu,1}$, the state $\rho=|v_{f}><v_{f}|$ becomes
\[
\rho\rightarrow\frac{P_{n+\mu,1}\rho P_{n+\mu,1}}{Tr\rho P_{n+\mu,1}%
}=:\overline{\rho}%
\]
Thus the solvability of the SAT problem is reduced to check that $\rho
^{\prime}\neq0$. The difficulty is that the probability
\[
Tr\overline{\rho}P_{n+\mu,1}=\Vert P_{n+\mu,1}\left\vert v_{f}\right\rangle
\Vert^{2}={\frac{|T(\mathcal{C}_{0})|}{2^{n}}}%
\]
is very small in some cases, where $|T(\mathcal{C}_{0})|$ is the cardinality
of the set $T(\mathcal{C}_{0})$, of all the truth functions $t$ such that
$t(\mathcal{C}_{0})=1.$

We put $q\equiv$\ $\sqrt{{\frac{r}{2^{n}}}}$ with $r\equiv|T(C_{0})|$ . Then
if $r$\ is suitably large to detect it, then the SAT problem is solved in
polynomial time. However, for small $r,$\ the probability is very small so
that we in fact do not get an information about the existence of the solution
of the equation $t(C_{0})=1,$\ hence in such a case we need further deliberation.

Let go back to the SAT algorithm. After the quantum computation, the quantum
computer will be in the state
\[
\left\vert v_{f}\right\rangle =\sqrt{1-q^{2}}\left\vert \varphi_{0}%
\right\rangle \otimes\left\vert 0\right\rangle +q\left\vert \varphi
_{1}\right\rangle \otimes\left\vert 1\right\rangle
\]
where $\left\vert \varphi_{1}\right\rangle $ and $\left\vert \varphi
_{0}\right\rangle $ are normalized $n$ (=$n+\mu)$ qubit states and
$q=\sqrt{r/2^{n}}.$ Effectively our problem is reduced to the following $1$
qubit problem: The above state $\left\vert v_{f}\right\rangle $ is reduced to
the state
\[
\left\vert \psi\right\rangle =\sqrt{1-q^{2}}\left\vert 0\right\rangle
+q\left\vert 1\right\rangle ,
\]
and we want to distinguish between the cases $q=0$ and $q>0$(small positive number).

\quad It will be not possible to amplify, by a unitary transformation, the
above small positive $q$ into suitable large one to be detected, e.g.,
$q>1/2,$with staying $q=0$ as it is. The amplification would be not possible
if we use the standard model of quantum computations with a unitary evolution.
What we did in \cite{OV1,OV2} is to propose to use the output $\left\vert
\psi\right\rangle $ of the quantum computer as an input for another device
involving chaotic dynamics. That is, it is proposed to combine quantum
computer with a chaotic dynamics amplifier in \cite{OV1,OV2}. Such a quantum
chaos computer is a new model of computations and we could demonstrate that
the amplification is possible in the polynomial time.

\subsection{Chaos algorithm of SAT}

Here we will argue that chaos can play a constructive role in computations
(see \cite{OV1,OV2} for the details). Chaotic behavior in a classical system
usually is considered as an exponential sensitivity to initial conditions. It
is this sensitivity we would like to use to distinguish between the cases
$q=0$ and $q>0$ mentioned in the previous section.

\quad Consider the so called logistic map
\[
x_{n+1}=ax_{n}(1-x_{n})\equiv g(x),~~~x_{n}\in\left[  0,1\right]  .
\]

\noindent\noindent\noindent The properties of the map depend on the parameter
$a.$ If we take, for example, $a=3.71,$ then the Lyapunov exponent is
positive, the trajectory is very sensitive to the initial value and one has
the chaotic behavior \cite{O2}. It is important to notice that if the initial
value $x_{0}=0,$ then $x_{n}=0$ for all $n.$

\quad It is known \cite{Deu} that any classical algorithm can be implemented
on quantum computer. Our quantum chaos computer will be consisting from two
blocks. One block is the ordinary quantum computer performing computations
with the output $\left\vert \psi\right\rangle =\sqrt{1-q^{2}}\left\vert
0\right\rangle +q\left\vert 1\right\rangle $. The second block is a computer
performing computations of the \textit{classical} logistic map. This two
blocks should be connected in such a way that the state $\left\vert
\psi\right\rangle $ first be transformed into the density matrix of the form
\[
\rho=q^{2}P_{1}+\left(  1-q^{2}\right)  P_{0}%
\]
where $P_{1}$ and $P_{0}$ are projectors to the state vectors $\left\vert
1\right\rangle $ and $\left\vert 0\right\rangle .$ This connection is in fact
nontrivial and actually it should be considered as the third block. One has to
notice that $P_{1}$ and $P_{0}$ generate an Abelian algebra which can be
considered as a classical system. In the second block the density matrix
$\rho$ above is interpreted as the initial data $\rho_{0}$, and we apply the
logistic map as
\[
\rho_{m}=\frac{(I+g^{m}(\rho_{0})\sigma_{3})}{2}%
\]
where $I$ is the identity matrix and $\sigma_{3}$ is the z-component of Pauli
matrix on $\mathbf{C}^{2}.$ \ To find a proper value $m$ we finally measure
the value of $\sigma_{3}$ in the state $\rho_{m}$ such that%

\[
M_{m}\equiv tr\rho_{m}\sigma_{3}.
\]
We obtain

\begin{theorem}
\quad%
\[
\rho_{m}=\frac{(I+g^{m}(q^{2})\sigma_{3})}{2},\text{ and }M_{m}=g^{m}(q^{2}).
\]

\end{theorem}

\quad Thus the question is whether we can find such a $m$ in polynomial steps
of $n\ $satisfying the inequality $M_{m}\geq\frac{1}{2}$ for very small but
non-zero $q^{2}.$ Here we have to remark that if one has $q=0$ then $\rho
_{0}=P_{0}$ and we obtain $M_{m}=0$ for all $m.$ If $q\neq0,$ the stochastic
dynamics leads to the amplification of the small magnitude $q$ in such a way
that it can be detected as is explained below. The transition from $\rho_{0}$
to $\rho_{m}$ is nonlinear and can be considered as a classical evolution
because our algebra generated by $P_{0}$ and $P_{1}$ is abelian. The
amplification can be done within at most 2n steps due to the following
propositions. Since $g^{m}(q^{2})$ is $x_{m}$ of the logistic map
$x_{m+1}=g(x_{m})$ with $x_{0}=q^{2},$ we use the notation $x_{m}$ in the
logistic map for simplicity.

\begin{theorem}
For the logistic map $x_{n+1}=ax_{n}\left(  1-x_{n}\right)  $ with $a$
$\in\left[  0,4\right]  $ and $x_{0}\in\left[  0,1\right]  ,$ let $x_{0}\ $be
$\frac{1}{2^{n}}$ and a set $J\ $be $\left\{  0,1,2,\cdots,n,\cdots2n\right\}
.$ If $a$ is $3.71,$ then there exists an integer $m$ in $J$ satisfying
$x_{m}>\frac{1}{2}.$
\end{theorem}

\begin{theorem}
Let $a$ and $n$ be the same in the above proposition. If there exists $m_{0}$
in $J$ such that $x_{m_{0}}>\frac{1}{2}$ $,$ then $m_{0}>\frac{n-1}{\log
_{2}3.71-1}.$
\end{theorem}

\quad According to these theorems, it is enough to check the value $x_{m}$
$(M_{m})$ around the above $m_{0}$ when $q$ is $\frac{1}{2^{n}}$ for a large
$n$. More generally, when $q$=$\frac{k}{2^{n}}$ with some integer $k,$ it is
similarly checked that the value $x_{m}$ $(M_{m})$ becomes over $\frac{1}{2}$
within at most 2n steps.

The complexity of the quantum algorithm for the SAT problem was discussed in
Section 3 to be in polynomial time. We have only to consider the number of
steps in the classical algorithm for the logistic map performed on quantum
computer. It is the probabilistic part of the construction and one has to
repeat computations several times to be able to distinguish the cases $q=0$
and $q>0.$ Thus it seems that the quantum chaos computer can solve the SAT
problem in polynomial time.

In conclusion of \cite{OV2}, the quantum chaos computer combines the ordinary
quantum computer with quantum chaotic dynamics amplifier. It may go beyond the
usual quantum Turing algorithm, but such a device can be powerful enough to
solve the \textbf{NP}-complete problems in the polynomial time. The detail
estimation of the complexity of the SAT algorithm is discussed in \cite{IA}.

In the next two sections we will discuss the SAT problem in a different view,
that is, we will show that the same amplification is possible by unitary
dynamics defined in the stochastic limit.

\section{ Quantum Adaptive Algorithm of SAT}

The idea to develop a mathematical approach to adaptive systems, i.e. those
systems whose properties are in part determined as responses to an environment
\cite{AI,KOT}, were born in connection with some problems of quantum
measurement theory and chaos dynamics.

The mathematical definition of adaptive system is in terms of observables,
namely: \textit{an adaptive system is a composite system whose }interaction
depends on a fixed observable (typically in a measurement process, this
observable is the observable one wants to measure). Such systems may be called
\textit{observable--adaptive.}

In the paper \cite{AO2} we extended this point of view by introducing another
natural class of adaptive systems which, in a certain sense, is the dual to
the above defined one, namely the class of \textit{state--adaptive} systems.
These are defined as follows: a state--adaptive system is a composite system
whose interaction depends on the state of at least one of the sub--systems at
the instant in which the interaction is switched on. We applied the
state-adaptivity to quantum computation.

The difference between state--adaptive systems and nonlinear dynamical systems
should be emphasized:

(i) in nonlinear dynamical systems (such as those whose evolution is described
by the Boltzmann equation, or nonlinear Schr\"{o}dinger equation, $\dots$, )
the interaction Hamiltonian depends on the state at each time $t$:
$H_{I}=H_{I}(\rho_{t})$ $;$ $\forall t$ .

(ii) in state--adaptive dynamical systems (such as those considered in the
present paper) the interaction Hamiltonian depends on the state only at time
$t=0$: $H_{I}=H_{I}(\rho_{0}).$

Now from the general theory of stochastic limit \cite{ALV} one knows that,
under general ergodicity conditions, an interaction with an environment drives
the system to a dynamical (but not necessarily thermodynamical) equilibrium
state which depends on the initial state of the environment and on the
interaction Hamiltonian.

Therefore, if one is able to realize experimentally these state dependent
Hamiltonians, one would be able to drive the system $S$ to a pre--assigned
dynamical equilibrium state depending on the input state $\psi_{0} $.

In the following subsection we will substantiate the general scheme described
above with an application to the SAT problem described in the previous sections.

\subsection{Stochastic Limit and SAT Problem}

We illustrate the general scheme described in the previous section in the
simplest case when the state space of the system is $\mathcal{H}_{S}%
\equiv\mathbf{C^{2}}$. We fix an orthonormal basis of $\mathcal{H}_{S}$ as
$\{e_{0},e_{1}\}.$

The unknown state (vector) of the system at time $t=0$
\[
\psi:=\sum_{\varepsilon\in\{0,1\}}\alpha_{\varepsilon}e_{\varepsilon}%
=\alpha_{0}e_{0}+\alpha_{1}e_{1}\ ;\ \ \Vert\psi\Vert=1.
\]
In Sec. 3, $\alpha_{1}$corresponds to $q$ and $e_{j}$ does to $\left\vert
j\right\rangle $ $\left(  j=0,1\right)  .$ This vector is taken as input and
defines the interaction Hamiltonian in an external field
\begin{align*}
H_{I}  & =\lambda|\psi\rangle\langle\psi|\otimes(A_{g}^{+}+A_{g})\\
& =\sum\lambda\alpha_{\varepsilon}\overline{\alpha}_{\varepsilon
}|e_{\varepsilon}\rangle\langle e_{\varepsilon^{\prime}}|\otimes(A_{g}%
^{+}+A_{g})
\end{align*}
where $\lambda$ is a small coupling constant. Here and in the following
summation over repeated indices is understood.

The free system Hamiltonian is taken to be diagonal in the $e_{\varepsilon}%
$--basis
\[
H_{S}:=\sum_{\varepsilon\in\{0,1\}}E_{\varepsilon}|e_{\varepsilon}%
\rangle\langle e_{\varepsilon}|=E_{0}|e_{0}\rangle\langle e_{0}|+E_{1}%
|e_{1}\rangle\langle e_{1}|
\]
and the energy levels are ordered so that $E_{0}<E_{1}.$ Thus there is a
single Bohr frequency $\omega_{0}:=E_{1}-E_{0}>0.$ The $1$--particle field
Hamiltonian is
\[
S_{t}g(k)=e^{it\omega(k)}g(k)
\]
where $\omega(k)$ is a function satisfying the basic analytical assumption of
the stochastic limit. Its second quantization is the free field evolution
\[
e^{itH_{0}}A_{g}e^{-itH_{0}}=A_{S_{t}g}%
\]
We can distinguish two cases as below, whose cases correspond to two cases of
Sec. 3, i.e., $q>0$ and $q=0.$

\textbf{Case (1)}. If $\alpha_{0},\alpha_{1}\not =0$ , then, according to the
general theory of stochastic limit (i.e., $t\rightarrow t/\lambda^{2})$
\cite{ALV}, the interaction Hamiltonian $H_{I}$ is in the same universality
class as
\[
\tilde{H}_{I}=D\otimes A_{g}^{+}+D^{+}\otimes A_{g}%
\]
where $D:=|e_{0}\rangle\langle e_{1}|.$ The interaction Hamiltonian at time
$t$ is then
\[
\tilde{H}_{I}(t)=e^{-it\omega_{0}}D\otimes A_{S_{t}g}^{+}%
+\hbox{ h.c.}=D\otimes A^{+}(e^{it(\omega(p)-\omega_{0})}g)+\hbox{ h.c.}
\]
and the white noise $\left(  \left\{  b_{t}\right\}  \right)  $ Hamiltonian
equation associated, via the stochastic golden rule, to this interaction
Hamiltonian is
\[
\partial_{t}U_{t}=i(Db_{t}^{+}+D^{+}b_{t})U_{t}%
\]
Its causally normal ordered form is equivalent to the stochastic differential
equation
\[
dU_{t}=(iDdB_{t}^{+}+iD^{+}dB_{t}-\gamma_{-}D^{+}Ddt)U_{t},
\]
where $dB_{t}:=b_{t}dt.$

The causally ordered inner Langevin equation is
\begin{align*}
dj_{t}(x)  & =dU_{t}^{\ast}xU_{t}+U_{t}^{\ast}xdU_{t}+dU_{t}^{\ast}xdU_{t}\\
& =U_{t}^{\ast}(-iD^{+}xdB_{t}-iDxdB_{t}^{+}-\overline{\gamma}_{-}%
D^{+}Dxdt+ixDdB_{t}^{+}\\
& +ixD^{+}dB_{t}-\gamma_{-}xD^{+}Ddt+\gamma_{-}D^{+}xDdt)U_{t}\\
& =ij_{t}([x,D^{+}])dB_{t}+ij_{t}([x,D])dB_{t}^{+}\\
& -(\hbox{Re }\gamma_{-})j_{t}(\{D^{+}D,x\})dt+i(Im\gamma_{-})j_{t}%
([D^{+}D,x])dt\\
& +j_{t}(D^{+}xD)(\hbox{Re }\gamma_{-})dt,
\end{align*}
where $j_{t}(x):=$ $U_{t}^{\ast}xU_{t}.$ Therefore the master equation is
\begin{align*}
{\frac{d}{dt}}\,P^{t}(x)  & =(Im\gamma)i[D^{+}D,P^{t}(x)]-(\hbox{Re}\gamma
_{-})\{D^{+}D,P^{t}(x)\}\\
& +(\hbox{Re }\gamma_{-})D^{+}P^{t}(x)D
\end{align*}
where $D^{+}D=|e_{1}\rangle\langle e_{1}|$ and $D^{+}xD=\langle e_{0}%
,xe_{0}\rangle|e_{1}\rangle\langle e_{1}|.$

The dual Markovian evolution $P_{\ast}^{t}$ acts on density matrices and its
generator is
\[
L_{\ast}\rho=(Im\gamma_{-})i[\rho,D^{+}D]-(\hbox{Re }\gamma_{-})\{\rho
,D^{+}D\}+(\hbox{Re }\gamma_{-})D\rho D^{+}%
\]
Thus, if $\rho_{0}=|e_{0}\rangle\langle e_{0}|$ one has
\[
L_{\ast}\rho_{0}=0
\]
so $\rho_{0}$ is an invariant measure. From the Frigerio--Fagnola--Rebolledo
criteria, it is the unique invariant measure and the semigroup $\exp(tL_{\ast
})$ converges exponentially to it.

\textbf{Case (2)}. If $\alpha_{1}=0,$ then the interaction Hamiltonian $H_{I}$
is
\[
H_{I}=\lambda|e_{0}\rangle\langle e_{0}|\otimes(A_{g}^{+}+A_{g})
\]
and, according to the general theory of stochastic limit, the reduced
evolution has no damping and corresponds to the pure Hamiltonian
\[
H_{S}+|e_{0}\rangle\langle e_{0}|=(E_{0}+1)|e_{0}\rangle\langle e_{0}%
|+E_{1}|e_{1}\rangle\langle e_{1}|
\]
therefore, if we choose the eigenvalues $E_{1},E_{0}$ to be integers (in
appropriate units), then the evolution will be periodic.

Since the eigenvalues $E_{1},E_{0}$ can be chosen a priori, by fixing the
system Hamiltonian $H_{S}$, it follows that the period of the evolution can be
known a priori. This gives a simple criterium for the solvability of the SAT
problem because, by waiting a sufficiently long time one can experimentally
detect the difference between a damping and an oscillating behavior.

A precise estimate of this time can be achieved either by theoretical methods
or by computer simulation. Both methods will be analyzed in the full paper
\cite{AO3}.

\begin{conclusion}
We pointed out that it is possible to distinguish two different states,
$\sqrt{1-q^{2}}\left\vert 0\right\rangle +q\left\vert 1\right\rangle $
$\left(  q\neq0\right)  $ and $\left\vert 0\right\rangle $ by means of the
adaptive dynamics with the stochastic limit.
\end{conclusion}

\begin{conclusion}
Finally we remark that our algorithm can be described by a deterministic
generalized quantum Turing machine \cite{IOV,AO3}.
\end{conclusion}

\section{Comparison of Various Quantum Mutual Type Entropies}

There exist several different types of quantum mutual entropy. The classical
mutual entropy was introduced by Shannon to discuss the information
transmission from an input system to an output system \cite{IKO}. Then
Kolmogorov \cite{Kol}, Gelfand and Yaglom \cite{GY} gave a measure theoretic
expression of the mutual entropy by means of the relative entropy defined by
Kullback and Leibler. The Shannon's expression of the mutual entropy is
generalized to one for finite dimensional quantum (matrix) case by Holevo
\cite{Hol,Ing}. Ohya took the measure theoretic expression by KGY and defined
quantum mutual entropy by means of quantum relative entropy \cite{O3,O4}.
Recently Shor \cite{Sho2} and Bennett et al \cite{BSST} took the coherent
information and defined the mutual type entropy to discuss a Shannon's coding
theorem. In this section, we compare these mutual types entropies.

The most gereral form of the quantum mutual entropy defined by Ohya,
generalizing the KGY measure theoretic mutual entropy, is given as
\[
I_{1}\left(  \varphi;\Lambda\right)  =\sup\left\{  \int_{\mathcal{S}}%
S^{AU}\left(  \Lambda\omega,\Lambda\varphi\right)  d\mu;\text{ }\mu\in
M_{\varphi}\left(  \mathcal{S}\right)  \right\}  .
\]
Here $\mathcal{S}$ is the set of all states in a certain C*-algebra (or von
Neumann algebra) describing a quantum system, $S^{AU}\left(  \cdot
,\cdot\right)  $ is the relative entropy of Araki \cite{Ara} or Uhlmann
\cite{Uhl} and $\mu$ is a measure decomposing the state $\varphi$ into
extremal orthogonal states, i.e., $\varphi=\int_{ex\mathcal{S}}\omega d\mu,$in
$\mathcal{S}$, whose set is denoted by $M_{\varphi}\left(  \mathcal{S}\right)
.$

In the case that the C*-algebra is $\mathbf{B}\left(  \mathcal{H}\right)  $
and $\mathcal{S}$ is the set of all density operators, the above mutual
entropy goes to%

\[
I_{1}\left(  \rho;\Lambda\right)  =\sup\left\{  \sum_{n}S^{U}\left(  \Lambda
E_{n},\Lambda\rho\right)  ;\text{ }\rho=\sum_{n}\lambda_{n}E_{n}\right\}  ,
\]
where $\rho$ is a density operator (state), $S^{U}\left(  \cdot,\cdot\right)
$ is Umegaki's relative entropy and $\rho=\sum_{n}\lambda_{n}E_{n}$ is a
Schatten-von Neumann (one dimensional spectral) decomposition. The SN
decomposition is not always unique unless $\mathcal{S}$ is Choque simplex, so
we take the supremum over all possible decompositions. It is easily shown that
we can take orthogonal decomposition instead of the SN decomposition \cite{O5}.

These quantum mutual entropy are completely quantum, namely, they describe the
information transmission from a quantum input to a quantum output. When the
input system is classical, the state $\rho$ is a probability distribution and
the Schatten-von Neumann decomposition is unique with delta measures
$\delta_{n}$ such that $\rho=\sum_{n}\lambda_{n}\delta_{n}.$ In this case we
need to code the classical state $\rho$ by a quantum state, whose process is a
quantum coding described by a channel $\Gamma$ such that $\Gamma\delta
_{n}=\sigma_{n}$ (quantum state) and $\sigma\equiv\Gamma\rho=\sum_{n}%
\lambda_{n}\sigma_{n}$. Then the quantum mutual entropy $I_{1}\left(
\rho;\Lambda\right)  $ becomes Holevo's one, that is,%

\[
I_{1}\left(  \rho;\Lambda\Gamma\right)  =S\left(  \Lambda\sigma\right)
-\sum_{n}\lambda_{n}S\left(  \Lambda\sigma_{n}\right)
\]
when $\sum_{n}\lambda_{n}S\left(  \Lambda\sigma_{n}\right)  $ is finite.

Let us discuss the entropy exchange \cite{BNS}. For a state $\rho$, a channel
$\Lambda$ is defined by an operator valued measure $\left\{  A_{j}\right\}  $
such as $\Lambda\left(  \cdot\right)  \equiv\sum_{j}A_{j}^{\ast}\cdot A_{j}%
.$Then define a matrix $W=\left(  W_{ij}\right)  $ with $W_{ij}=\frac
{trA_{i}^{\ast}\rho A_{j}}{tr\Lambda\rho},$by which the entropy exchange is
defined by%

\[
S_{e}\left(  \rho,\Lambda\right)  =-trW\log W.
\]

Using the above entropy exchange, two mutual type entropies are defined as
below and they are applied to the study of quantum version of Shannon' coding
theorem \cite{BNS,Sho2,BSST}. The first one is called the coherent information
$I_{2}\left(  \rho;\Lambda\right)  $ and the second one is $I_{3}\left(
\rho,\Lambda\right)  $, which are defined by%

\begin{align*}
I_{2}\left(  \rho,\Lambda\right)   & \equiv S\left(  \Lambda\rho\right)
-S_{e}\left(  \rho,\Lambda\right)  ,\\
I_{3}\left(  \rho,\Lambda\right)   & \equiv S\left(  \rho\right)  +S\left(
\Lambda\rho\right)  -S_{e}\left(  \rho,\Lambda\right)  .
\end{align*}

By comparing these mutual entropies for information communication processes,
we have the following theorem \cite{OW}:

\begin{theorem}
When $\left\{  A_{j}\right\}  $ is a projection valued measure and
dim(ran$A_{j})=1,$ for arbitary state $\rho$ we have (1) $I_{1}\left(
\rho,\Lambda\right)  \leq\min\left\{  S\left(  \rho\right)  ,S\left(
\Lambda\rho\right)  \right\}  $, (2) $I_{2}\left(  \rho,\Lambda\right)  =0,$
(3) $I_{3}\left(  \rho,\Lambda\right)  =S\left(  \rho\right)  .$
\end{theorem}

From this theorem, the entropy $I_{1}\left(  \rho,\Lambda\right)  $ only
satisfies the inequality held in classical systems, so that only this entropy
can be a candidate as quantum extension of the classical mutual entropy. Other
two entropies can describe a sort of entanglement between input and output,
such a correlation can be also described by quasi-mutual entropy, a slight
generalization of $I_{1}\left(  \rho,\Lambda\right)  ,$discussed in
\cite{O5,BO}.

\begin{acknowledgement}
The author thanks IIAS and SCAT for financial support of this work.
\end{acknowledgement}


\begin{thebibliography}{99}                                                                                               %
\bibitem {AI}{\small L. Accardi and K. Imafuku: Control of Quantum States by
Decoherence, to appear in Open Systems and Information Dynamics, 2003}

\bibitem {ALV}{\small L.Accardi, Y.G. Lu, I.Volovich: Quantum Theory and its
Stochastic Limit. Springer Verlag 2002; Japanese translation, Tokyo--Springer
2003.}

\bibitem {AO}{\small L.Accardi and M.Ohya, Compound channels, transition
expectations, and liftings, Appl. Math. Optim., Vol.39, 33-59, 1999.}

\bibitem {AO2}{\small L.Accardi and M.Ohya, A stochastic limit approach to the
SAT problem, to appear.}

\bibitem {AO3}{\small L.Accardi and M.Ohya, Generalized quantum Turing machine
and stochastic limit for the SAT problem, in preparation.}

\bibitem {Ara}{\small \ H.Araki: Relative entropy of states of von Neumann
Algebras, Publ.RIMS, Kyoto Univ.Vol.11, 809-833, (1976); Relative entropy for
states of von Neumann algebras II, Publ.RIMS, Kyoto Univ., 13, pp.173--192,
(1977)}

\bibitem {AS}{\small L.Accardi and Ruben Sabbadini, On the Ohya--Masuda
quantum SAT Algorithm, Proceedings Intern.Conf. \textquotedblright
Unconventional Models of Computations\textquotedblright, I. Antoniou, C.S.
Calude, M. Dinneen (eds.) Springer 2001 ; Preprint Volterra, N. 432, 2000}

\bibitem {BNS}{\small H.Barnum, M.A.Nielsen and B.W.Schumacher, Information
transmission through a noisy quantum channel, Physical Review A, Vol.57, No.6,
4153-4175, 1998.}

\bibitem {BO}{\small V.P.Belavkin and M.Ohya, Quantum entropy and information
in discrete entangled states, Infinite Dimensional Analysis, Quantum
Probability and Related Topics, Vol.4, No.2, 137-160 (2001);Quantum
entanglements and entangled mutual entropy, Proc.R.Soc.Lond.A.458, 209-231
(2002)}

\bibitem {BSST}{\small C.H. Bennett , P.W. Shor, J.A. Smolin , and A.V.
Thapliyalz, Entanglement-Assisted Capacity of a Quantum Channel and the
Reverse Shannon Theorem, quant-ph/0106052.}

\bibitem {BV}{\small E.Bernstein and U.Vazirani, Quantum complexity theory,
Proc.of the 25th Annual ACM Symposium on Theory of Computing, ACM, New York,
pp.11-22.(1993), SIAM\ Journal\ on\ Computing\ 26,\ 1411 (1997)\ }

\bibitem {BBBV}{\small C. H. Bennett, E. Bernstein, G. Brassard and U.
Vazirani, Strengths and Weaknesses of Quantum Computing, quant-ph/9701001.}

\bibitem {Cle}{\small R. Cleve, An Introduction to Quantum Complexity Theory,
quant-ph/9906111.}

\bibitem {Deu}{\small D. Deutsch, Quantum theory, the Church-Turing principle
and the universal quantum computer, Proc. of Royal Society of London series A,
400, pp.97-117, 1985.}

\bibitem {EJ}{\small A. Ekert and R. Jozsa, Quantum computation and Shor's
factoring algorithm, Reviews of Modern Physics, 68 No.3,pp.733-753, 1996.}

\bibitem {GJ}{\small M. Garey and D. Johnson, Computers and Intractability - a
guide to the theory of NP-completeness, Freeman, 1979.}

\bibitem {GY}{\small \ I.M.Gelfand and A.M.Yaglom, Calculation of the amount
of information about a random function contained in another such function,
Amer.Math.Soc.Transl., 12, pp.199-246, (1959)}

\bibitem {Hol}{\small \ A.S.Holevo, Some estimates for the amount of
information transmittable by a quantum communication channel (in Russian),
Problemy Peredachi Informacii, 9, pp.3-11, (1973)}

\bibitem {IA}{\small S. Iriyama and S. Akashi,\ Complexity of
Ohya-Masuda-Volovich algorithm, to appear}

\bibitem {IKO}{\small R.S.Ingarden, A.Kossakowski and M.Ohya, Information
Dynamics and Open Systems, Kluwer, (1997)}

\bibitem {IOV}{\small S. Iriyama, M. Ohya and I. Volovich, Generalized quantum
Turing machine and its application to the SAT chaos algorithm, TUS(Tokyo
University of Science) preprint, 2003.}

\bibitem {Ing}{\small R.S.Ingarden, Quantum information theory, Rep. Math.
Phys., 10, pp.43-73, 1976.}

\bibitem {Neu}{\small J.von Neumann, Die Mathematischen Grundlagen der
Quantenmechanik, Springer-Berlin, 1932.}

\bibitem {Kol}{\small A.N.Kolmogorov, Theory of transmission of information,
Amer. Math. Soc. Translation, Ser.2, 33, pp.291--321, 1963.}

\bibitem {KOT}{\small A.Kossakowski, M.Ohya and Y.Togawa, How can we observe
and describe chaos?, Open System and Information Dynamics 10(3): 221-233,
2003}

\bibitem {O2}{\small M. Ohya, Complexities and Their Applications to
Characterization of Chaos, Int. Journ. of Theort. Phy., 37, 495, 1998.}

\bibitem {O3}{\small M. Ohya, On compound state and mutual information in
quantum information theory, IEEE Trans.Information Theory, 29, pp.770--777
(1983)}

\bibitem {O4}{\small M. Ohya, Some aspects of quantum information theory and
their applications to irreversible processes, Rep.Math.Phys., Vol.27, 19-47,
(1989)}

\bibitem {O5}{\small M.Ohya, Fundamentals of quantum mutual entropy and
capacity, Open Systems and Information Dynamics, 6. No.1, 69-78, 1999.}

\bibitem {OM}{\small M.Ohya and N.Masuda, NP problem in Quantum Algorithm,
Open Systems and Information Dynamics, 7 No.1 33-39, 2000.}

\bibitem {OP}{\small M. Ohya and D.Petz, Quantum Entropy and its Use,
Springer-Verlag, (1993)}

\bibitem {OV1}{\small M.Ohya and I.Volovich, Quantum computing, NP-complete
problems and chaotic dynamics, Quantum Information II, eds. T.Hida and
K.Saito, World Sci. 2000; quant-ph/9912100 and J.Opt.B, 5,No.6 639-642, 2003}

\bibitem {OV2}{\small M. Ohya and I. Volovich, New quantum algorithm for
studying NP-complete problems, Rep.Math.Phys.,52, No.1,25-33, 2003}

\bibitem {OV3}{\small M.Ohya and I. Volovich, Quantum Information,
Computation, Cryptography and Teleportation, Springer-Verlag (to appear).}

\bibitem {OW}{\small M. Ohya and N.Watanabe, Remarks on quantum mutual
entropy, TUS preprint.}

\bibitem {PM}{\small D. Petz and M. Mosonyi, Stationary quantum source coding,
Journal of Mathematical Physics, Vol.42, 4857-4864, 2001.}

\bibitem {Sho}{\small P. Shor, Algorithm for quantum computation, Discrete
logarithm and factoring algorithm, Proceedings of the 35th Annual IEEE
Symposium on Foundation of Computer Science, pp.124-134, 1994.}

\bibitem {Sho2}{\small P.Shor, The quantum channel capacity and coherent
information, Lecture Notes, MSRI Workshop on Quantum Computation, 2002.}

\bibitem {Uhl}{\small A.Uhlmann, Relative entropy and the
Wigner-Yanase-Dyson-Lieb concavity in interpolation theory, Commun. Math.
Phys., Vol.54, 21-32, 1977.}

\bibitem {Um}{\small H.Umegaki, Conditional expectations in an operator
algebra IV(entropy and information), Kodai Math.Sem.Rep., Vol.14, 59-85,
(1962)}
\end{thebibliography}
\end{document}